\documentclass[a4paper,11pt]{article}
\usepackage[T1]{fontenc}
\usepackage[utf8]{inputenc}
\usepackage{lmodern}
\usepackage{graphicx}
\usepackage{epstopdf} 
\usepackage{color}
\usepackage{bbold}
\usepackage{amsmath}
\newcommand{\wt}{\widetilde}

\title{Gribov Problem in the Eletroweak and Gluon Confined Theories}

\author{R. L. P. G. Amaral$^{a}$\footnote{email: rubens@if.uff.br} ,
	V. E. R. Lemes$^{b}$\footnote{email: verlemes@gmail.com},
	O. S. Ventura$^{c}$\footnote{email: ozemar.ventura@cefet-rj.br}, L.C.Q.Vilar $^{b}$\footnote{email: lcqvilar@gmail.com}  \\
	\small \em $^a$Instituto de F\'{\i}sica, Universidade Federal do Fluminense\\
	\small \em Av. Litor\^anea S/N, Boa Viagem, Niter\'oi-RJ CEP. 24210-340,
	Brazil\\
	\small \em $^b$Instituto de F\'\i sica, Universidade do Estado do Rio de
	Janeiro,\\
	\small \em Rua S\~{a}o Francisco Xavier 524, Maracan\~{a}, Rio de Janeiro - RJ,
	20550-013, Brazil\\
	\small \em $^c$Departamento de F\'\i sica, Centro Federal de Educa\c{c}\~ao Tecnol\'ogica do Rio de
	Janeiro\\
	\small\em Av.Maracan\~a 249, 20271-110, Rio de Janeiro - RJ, Brazil}

\begin{document}
	\maketitle
	\begin{abstract}	
	We show that gauge fields after a spontaneous symmetry breaking (SSB) mechanism do not have a Gribov problem along the broken directions. In order to make this proof, we describe a gauge fixing procedure inspired on Morse´s theory leading to the concept of a gauge fixing generating functional. This approach is specially suited in order to understand the unitary gauges of ´t Hooft. The conclusion is that after a SSB process, the generalized Faddeev-Popov operator acquires a positive definite value along the broken directions. We show how this works in the eletroweak $SU(2)XU(1)$ theory, where such a result is in fact expected as in this case the gauge fields acquire masses after the SSB. Then we apply this development to the $SL(3,c)$ confining model of $\cite{ALVV20}$. The final result explains why such confining mechanism is actually a solution to Gribov problem for the confined gauge fields. These cases show that although confinement is not a general effect of the solution of the Gribov problem, as we can infer from the eletroweak example, its solution indeed seems to be necessary in order to achieve confinement. In the end, these conclusions allow us to speculate that the strong interaction confinement may be the result of some hidden SSB mechanism yet to be described.
		
\end{abstract}
	
\section{Introduction}

We know that a nonabelian gauge field theory has the so called Gribov problem \cite{Gribov}. This problem is based on the simple observation that in order to extract physically meaningful information from the euclidean partition function

\begin{eqnarray}
Z(j) = \int DA e^{-\int d^{4}x_{E} ( \frac{1}{4}F^{A}_{\mu\nu}F^{A}_{\mu\nu} - J^{A}_{\mu}A^{A}_{\mu})}
\label{pf}
\end{eqnarray}
one should take care with the measure. The gauge symmetry implies that several physically redundant and uncountable copies of any given gauge field configuration are being integrated over in ($\ref{pf}$), unless a gauge fixing procedure is built ensuring that just one exemplar is selected among those copies. One such procedure is for example the imposition of the Landau gauge condition

\begin{eqnarray}
\partial_{\mu}A^{A}_{\mu} =0 \,  
\label{lg}
\end{eqnarray}
which can be implemented in ($\ref{pf}$) through the Faddeev-Popov ansatz. The question then is whether once we generate a gauge copy of a field configuration specified by ($\ref{lg}$),

\begin{eqnarray}
 A^{A}_{\mu} & \rightarrow & \tilde{A}^{A}_{\mu} =  A^{A}_{\mu} + \delta A^{A}_{\mu}  \, , \nonumber \\
\delta A^{A}_{\mu}&=& -\left(\partial_{\mu}\epsilon^{A}+gf^{ABC}A^{B}_{\mu}\epsilon^{C}\right)= -(D_{\mu}\epsilon)^{A} \, ,
\label{dA}
\end{eqnarray}
is it possible that this copy could satisfy 

\begin{eqnarray}
\partial_{\mu} \tilde{A}^{A}_{\mu} =0  \implies \partial_{\mu} D_{\mu}\epsilon^{A} =0   \, , 
\label{ltilde}
\end{eqnarray}
the same gauge condition as ($\ref{lg}$)? If this happens, the gauge is not completely fixed, and the ambiguity in ($\ref{pf}$) still remains. Gribov explicitly showed that this is the case, and this became known as the Gribov problem. Put in another way, from ($\ref{ltilde}$) we say that the Landau gauge implies that the Faddeev-Popov operator admits null eigenvalues, and this leads to the ambiguity.

This problem has a rereading in terms of Morse theory, which is the study of topological invariants of any given manifold in terms of critical points of functions defined on it \cite{Milnor,Matsumoto}. This interpretation was given for the first time in \cite{Baal}, explored in \cite{LLVV14}, but actually its use may be traced back to the original construction made by Gribov \cite{Gribov}. Here we will summarize the more detailed presentation given in \cite{LLVV14}. It begins by seeing a gauge orbit of a given gauge configuration as a foliation of the whole gauge configuration space. To each of these gauge orbits we will give an appropriate topography, written as a functional generator $O$. We can choose as a topography, for example, the Hilbert norm of the gauge field

\begin{eqnarray}
O&=& \int d^{4}x\, \frac{1}{2} A^{A}_{\mu} A^{A}_{\mu} \, .
\label{Ol}
\end{eqnarray}
An infinitesimal displacement along the gauge orbit is determined by the gauge transformation ($\ref{dA}$), which in quantum mechanical terms is described by the set of nilpotent BRST transformations

\begin{eqnarray}
sA^{A}_{\mu}&=& -\left(\partial_{\mu}c^{A}+gf^{ABC}A^{B}_{\mu}c^{C}\right)= -(D_{\mu}c)^{A} \, , \nonumber \\
sc^{A}&=& \frac{g}{2}f^{ABC}c^{B}c^{C} \, , \nonumber \\
s\overline{c}^{A} &=& ib^{A} \, ,
\label{BRS}
\end{eqnarray}
with $c^{A}$ and $b^{A}$ the usual ghost field and Nakanishi-Lautrup multiplier respectively. Then, the variation of the gauge generating functional ($\ref{Ol}$) is given by

\begin{eqnarray}
sO &=&  \int d^{4}x\, c^{A}\partial_{\mu}A^{A}_{\mu} \, ,
\label{sO}
\end{eqnarray}
and the demand that a given configuration be a critical point in its orbit, $sO=0$, is equivalent to

\begin{eqnarray}
 \frac{\boldsymbol{\delta} sO}{\boldsymbol{\delta} c^{A}}= \partial_{\mu}A^{A}_{\mu}=0 \, .
\label{dO}
\end{eqnarray}
This is the Landau gauge condition ($\ref{lg}$), now understood as a critical point of ($\ref{Ol}$). At this point, one can implement this condition at the quantum action by adding the Faddeev-Popov gauge fixing sector

\begin{eqnarray}
S_{gf}&=& \int d^{4}x\left(  i b^{A}\partial_{\mu}A^{A}_{\mu} +  \overline{c}^{A}\partial_{\mu}(D_{\mu}c)^{A}\right) \, .
\label{Sgf}
\end{eqnarray}
Using the BRST language, we can translate this mechanism as an operation on the functional ($\ref{Ol}$)
\begin{eqnarray}
S_{gf}&=& s \int d^{4}x \overline{c}^{A}\frac{\boldsymbol{\delta} sO}{\boldsymbol{\delta} c^{A}}   \, ,
\label{SgfO}
\end{eqnarray}
which will be applied later to a generalized Morse functional. It is also useful to write this expression in its expanded form
\begin{eqnarray}
S_{gf}&=&  \int d^{4}x\, \left\{ i{b}^{A}\left(\frac{\boldsymbol{\delta} sO}{\boldsymbol{\delta} c^{A}}\right) - \overline{c}^{A}s\left(\frac{\boldsymbol{\delta} sO}{\boldsymbol{\delta} c^{A}}\right) \right\}  \, .
\label{SgfOe}
\end{eqnarray}

Another fundamental feature of the Landau gauge is that it satisfies the symmetry equation 

\begin{eqnarray}
\int  d^{4}x \left(\frac{\boldsymbol{\delta} S_{gf}}{\boldsymbol{\delta} c^{A}} -ig f^{ABC} \overline{c}^{B} \frac{\boldsymbol{\delta} S_{gf}}{\boldsymbol{\delta} b^{C}}\right) =0 \, .
\label{geq}
\end{eqnarray}
The importance of this anti-ghost equation is
that it is valid at the quantum level as it satisfies the Quantum Action Principle (QAP). Then it plays a major role in the study of the renormalization of non-abelian gauge theories (see \cite{Sor95} for a detailed introduction in BRST renormalization).

Returning to the Gribov ambiguity, the expression ($\ref{ltilde}$) pointing out the possible existence of copies satisfying Landau condition can now be translated as a second variation of the functional ($\ref{Ol}$) at the critical point ($\ref{dO}$),

\begin{eqnarray}
 s \frac{\boldsymbol{\delta} sO}{\boldsymbol{\delta} c^{A}} = 0 \hspace*{0,5 cm}  at \hspace*{0,5 cm}  \frac{\boldsymbol{\delta} sO}{\boldsymbol{\delta} c^{A}} = 0 \, .
\label{ddO}
\end{eqnarray}
In fact, this condition is implicit in the gauge fixing action ($\ref{Sgf}$), as it is just the on-shell equation of the anti-ghost $\overline{c}^{A}$ when $A_{\mu}$ satisfies  ($\ref{lg}$)

\begin{eqnarray}
 \partial^{2}c^{A} + g f^{ABC} A^{B}_{\mu}\partial^{\mu}c^{C} = 0.
 \label{gribeq}
\end{eqnarray}
It can also be read directly from ($\ref{SgfOe}$) as the Morse functional ($\ref{Ol}$) does not depend on $\overline{c}^{A}$.
Then,  ($\ref{ddO}$) expresses, in a generalized Morse functional form, the action on the ghost of the Faddev-Popov operator. We can now understand in simple terms the nature of the Gribov ambiguity. The imposition of a null second variation at a critical point does not fix its nature. Otherwise, if one is able to implement at the quantum action a condition as

\begin{eqnarray}
 \partial^{2}c^{A} + g f^{ABC} A^{B}_{\mu}\partial^{\mu}c^{C}   > 0 \, ,
\label{gcond}
\end{eqnarray}
it would be assured that the theory would be localized at a minimum of   ($\ref{Ol}$), as the Hessian of the Morse functional $O$ would have only positive definite eigenvalues. In fact, this is the characterization of Gribov's first region, and the implementation of this condition on a gauge theory has been the subject of several works (see the review \cite{Zw12} for an account of this extensive literature). Although many breakthroughs were obtained along this research, this implementation is still an open question. Anyway, it is a general belief that the correct description of the non-perturbative region, and consequently of confinement, requires the solution of this problem.

Our strategy in order to obtain a relation as  ($\ref{gcond}$) is to modify the functional $O$ in such a way as to arrive at a different gauge fixing through ($\ref{SgfO}$). This will lead us to a new gauge condition and an equation of motion for $\overline{c}^{A}$, that will be interpreted as a new Gribov equation. Our hope is that we may define the physical conditions to meet ($\ref{gcond}$) in the end.

 In the sequence of this work, we will study other generating functionals instead of ($\ref{Ol}$). In Section 2, for example, we will start from a functional that will lead us to a gauge fixing displaying characteristics of Feynman and Curcci-Ferrari gauges. We will show how the anti-ghost equation again satisfies the QAP but the Gribov ambiguity remains. Section 3 starts from a functional involving scalar fields and then an SSB phase can be reached. In this phase, this functional will describe a 't Hooft gauge, and we will see how unitarity implies that the gauge fields in the broken directions get localized in Gribov's first region. The example of a $SU(2)XU(1)$ theory is the subject of Section 4. We will show how the Gribov ambiguity is solved along the broken directions. In Section 5 we approach the $SL(3,c)$ complex gauge field theory of $\cite{ALVV20}$. Once more, the Gribov ambiguity is solved for the complex gauge fields along the broken directions, which in this case are related to i-particles and the formation of gluon condensates. In Section 6 we present our conclusions, stressing that the solution to the Gribov ambiguity may be found on a broken phase of non-abelian gauge theories. This may be a signal that the confinement for the strong interactions is associated to a spontaneously broken symmetry.

\section{Generalizing the Morse Funtional}

Before presenting an example of a generalized Morse functional, let us describe the situation with the Feynman gauge. As demanded by the standard procedure, this gauge fixing can be presented as a BRST trivial cocycle

\begin{eqnarray}
S_{gfF}&=& s \int d^{4}x \left(  \overline{c}^{A}\partial_{\mu}A^{A}_{\mu} +  \frac{ia}{2} b^{A}\overline{c}^{A}\right) \, .
\label{sgf}
\end{eqnarray}
The gauge condition that results from ($\ref{sgf}$)
\begin{eqnarray}
\partial_{\mu}A^{A}_{\mu} -ab^{A} =0 \, , 
\label{gcf}
\end{eqnarray}
leads to the Feynman gauge propagator, but we can easily see that it is not helpful in the equation of motion of the anti-ghost $\overline{c}^{A}$ in order to arrive at a meaningful Gribov equation. This is not surprising, as the expression in ($\ref{sgf}$) cannot be directly obtained from a functional form as ($\ref{SgfO}$).

On the other hand, we find $S_{gfF}$ as part of a gauge fixing that is generated by the functional
\begin{eqnarray}
O_{2}&=&  \int d^{4}x \left( \frac{1}{2} A^{A}_{\mu} A^{A}_{\mu} + \frac{a}{2} \overline{c}^{A}c^{A}\right) \, .
\label{O2}
\end{eqnarray}
This is the first generalized functional that we want to study. Our interest begins by noticing that once it depends explicitly on $\overline{c}^{A}$, the gauge condition will obey the general pattern

\begin{align}
\frac{\boldsymbol{\delta} sO_{2}}{\boldsymbol{\delta} c^{A}} = &-\overline{c}^{B}  \frac{\boldsymbol{\delta}}{\boldsymbol{\delta} b^{A}} \left\{ s\left(\frac{\boldsymbol{\delta} sO_{2}}{\boldsymbol{\delta} c^{B}}\right) \right\}- b^{B}  \frac{\boldsymbol{\delta}}{\boldsymbol{\delta} b^{A}} \left(\frac{\boldsymbol{\delta} sO_{2}}{\boldsymbol{\delta} c^{B}}\right)  \, ,
\label{gcg}
\end{align}
where a non null contribution to the right is expected. This means that this gauge condition now does not fix the gauge at  critical points of $O_{2}$.  Continuing with the calculations, the BRST variation of ($\ref{O2}$)  gives
\begin{eqnarray}
sO_{2}&=& \int d^{4}x \left(  c^{A}\partial_{\mu}A^{A}_{\mu} +  \frac{ia}{2} b^{A}c^{A} - \frac{ag}{4} f^{ABC} \overline{c}^{A} c^{B} c^{C}\right) \, ,
\label{sO2}
\end{eqnarray}
and following the recipe given in ($\ref{SgfO}$), we find the gauge fixing action related to this functional

\begin{eqnarray}
S_{gf}&=&  \int d^{4}x \left(  i b^{A}\partial_{\mu}A^{A}_{\mu} -  \frac{a}{2}  b^{A}b^{A} - iagf^{ABC} b^{A} \overline{c}^{B} c^{C}\right.\nonumber \\
&-& \left.a\frac{g^{2}}{4}f^{ABC}f^{CDE} \overline{c}^{A} \overline{c}^{B} c^{D} c^{E} + \overline{c}^{A}\partial_{\mu}(D_{\mu}c)^{A}\right)
\label{gfO2}
\end{eqnarray}
It is straightforward to see that we obtain  a Feynman gauge propagator once we integrate on the multipliers $b^{A}$. Bur the four ghost interaction resembles the gauge fixing introduced in  $\cite{CF}$, the Curci-Ferrari model. The novelty is that here we can derive the same anti-ghost equation of ($\ref{geq}$), showing that in this gauge we can control the renormalization of the ghost field. Also, as anticipated in  ($\ref{gcg}$), the gauge condition, obtained from the equation of the multiplier ${b}^{A}$ in  ($\ref{gfO2}$)

\begin{eqnarray}
\partial_{\mu}A^{A}_{\mu} +ia b^{A} - agf^{ABC}  \overline{c}^{B} c^{C}=0 \, ,
\label{gcO2}
\end{eqnarray}
does not fix a critical point of the functional ($\ref{O2}$). Nevertheless, ($\ref{gcO2}$) is  a critical point of another functional, as can be easily guessed

\begin{eqnarray}
O_{g}&=&  \int d^{4}x \left( \frac{1}{2} A^{A}_{\mu} A^{A}_{\mu} + a \overline{c}^{A}c^{A}\right) \, .
\label{Og}
\end{eqnarray}
Then we see that this gauge fixing method is still associated to a minimization process, although not of its generating functional. Another interesting point is that as we are dealing with a multiplier, we can fix   ${b}^{A}$ on-shell as

\begin{eqnarray}
 b^{A} = -igf^{ABC}  \overline{c}^{B} c^{C} \, ,
\label{b}
\end{eqnarray}
and from ($\ref{gcO2}$) we find that the gauge configurations will again satisfy the Landau condition. In this way, this process is meeting the critical points of ($\ref{Og}$) and of ($\ref{Ol}$) simultaneously. This is a novelty allowed by this gauge fixing method.

Now we follow to the equation of motion of the anti-ghost. Again, in general terms, from the dependence of ($\ref{O2}$) on $\overline{c}^{A}$ we obtain

\begin{eqnarray}
   s\left(\frac{\boldsymbol{\delta} sO_{2}}{\boldsymbol{\delta} c^{A}}\right)   = -\int_y{\overline c}^{B}_y s\left(\frac{\boldsymbol{\delta}}{\boldsymbol{\delta} \overline{c}^{A}} \frac{\boldsymbol{\delta} sO_{2}}{\boldsymbol{\delta} c^{B}_y}\right) +i\int_y{b}^{B}_y \frac{\boldsymbol{\delta}}{\boldsymbol{\delta} \overline{c}^{A}} \frac{\boldsymbol{\delta} sO_{2}}{\boldsymbol{\delta} c^{B}_y}  \, .
\label{eagg}
\end{eqnarray}
Compared to the Landau case, eq.($\ref{ddO}$), we see that, in principle, the Gribov  equation gets changed once we impose the evaluation of ($\ref{eagg}$) at the critical point ($\ref{gcg}$). There is a hope to meet a condition as ($\ref{gcond}$), but once we use the condition ($\ref{gcO2}$) on ($\ref{eagg}$), an unexpected cancellation happens and we find again the same Gribov ambiguity expressed in ($\ref{gribeq}$). This means that our method of constructing the gauge fixing from the functional $O_{2}$ in ($\ref{O2}$), using the gauge condition ($\ref{b}$), is actually fixing the gauge in the same critical point of ($\ref{Ol}$), with the same ambiguity of the Landau gauge. 

Then, the generalization to the functional ($\ref{O2}$) does not bring us any closer to Gribov's first region. Actually, we could have expected this conclusion, as this gauge fixing is not associated to a different physical condition, it is just a mathematical generalization of the Landau case. Possibly we should look for this answer in the gauge fixing of a distinct physical context. Anyway, this exercise just done will be useful, as we will see.

\section{Morse Functional of a Spontaneous Symmetry Broken Phase}

If we want to talk about a SSB process, we must introduce a scalar field $\varphi^{A}$ transforming as
\begin{eqnarray}
s\varphi^{A}&=& gf^{ABC}c^{B}\varphi^{C} \, . 
\label{sphi}
\end{eqnarray}
Obviously, as the scalar mass term is invariant under ($\ref{sphi}$),
\begin{eqnarray}
s \int d^{4}x \left(\varphi^{A}\varphi^{A}\right)=0 \, ,
\label{sphi2}
\end{eqnarray}
adding such a term to the generalized functional ($\ref{O2}$), in principle, would not make any change in the conclusions we arrived at the last Section. But everything changes if we move from the symmetric vacuum, implicit when we write ($\ref{sphi}$), to the vacuum of the broken phase (from now on, we follow the description and notations of \cite{ALVV22}). At this phase, the scalar field acquires a non-null vacuum expectation value 
\begin{equation}
 \varphi^{A}=\chi^{A}+\mu\delta^{Az} \, , 
 \label{vev0}
\end{equation}
where $z$ is the direction of the breaking. The shift in $\varphi$ implies $<\chi>=0$. In the internal space the directions that commute with $z$ are designated by lowercase letters from the middle of the alphabet, as i,j,k... These symmetric directions of the new vacuum may form subgroups of the original symmetry group $G$. Then, in some equations, we will use a subindex $(N)$ to refer to $N$ possible distinct subgroups. For example $A^{i(N)}$ will refer to the $i$ component of the gauge field associated to the subgroup of label $N$. We also have the broken directions, which  do not commute with $z$. They are designated by lowercase letters from the beginning of the alphabet, as a,b,c... In \cite{ALVV22}, the algebraic result of the broken theory is specified  by saying that the structure constants of this phase to have non-null contributions are only
\begin{eqnarray}
&&f^{zab},\hspace{3mm}f^{ijk},\hspace{3mm}f^{iab} \, .
\label{fs}
\end{eqnarray}
This is the Cartan decomposition of a Lie algebra for a symmetric space when $f^{abc}$ vanishes \cite{Wei96}. This may not be the most general possibility, but this simplification is already sufficiently general for our purposes.

As discussed in  \cite{ALVV22}, in the broken phase, physical observables are characterized by a new set of nilpotent BRST transformations that were called $s_{q}$

\begin{align}
s_{q}A^{i(N)}_{\mu}=& -\partial_{\mu}c^{i(N)}- g_{N}f^{ijk(N)}A^{j(N)}_{\mu}c^{k(N)}\,,&s_{q}A^{z}_{\mu}= &-\partial_{\mu}c^{z}  , \nonumber \\
s_{q}A^{a}_{\mu}=& - \sum_{N} g_{N}f^{abi(N)}A^{b}_{\mu}c^{i(N)} - g^{\prime} f^{abz}A^{b}_{\mu}c^{z} \, ,&& \nonumber \\
s_{q}\chi^{i(N)}=&g_{N}f^{ijk(N)}c^{j(N)}\chi^{k(N)} \, , & s_{q}\chi^{z}=&0 \, , \nonumber  \\
s_{q}\chi^{a}=&  \sum_{N} g_{N}f^{aib(N)}c^{i(N)}\chi^{b} + g^{\prime} f^{azb}c^{z}\chi^{b} \, , &&\nonumber \\
s_{q}c^{i(N)}=&\frac{1}{2} g_{N}f^{ijk(N)}c^{j(N)}c^{k(N)} \, ,&
s_{q}c^{z}=&0 \, , \nonumber \\
s_{q}c^{a}=& \sum_{N} g_{N}f^{abi(N)}c^{b} c^{i(N)} + g^{\prime} f^{abz}c^{b} c^{z} \, .&&
\label{sq}
\end{align}
As has been done in \cite{ALVV22}, the  operator $\delta$ must be introduced such that the nilpotent operator $s_q+\delta$ allows the definition of a physically sensible gauge fixing for the broken phase. Here the following simplified version of  the operator $\delta$ is enough to discuss the Gribov problem

\begin{eqnarray}
\delta A^{a}_{\mu}&=& - \sum_{N} g_{N}f^{aib(N)}A^{i(N)}_{\mu}c^{b} -  g^{\prime} f^{azb}A^{z}_{\mu}c^{b} -\partial_{\mu}c^{a} \, , \nonumber \\
\delta \chi^{a}&=& \mu f^{abz}c^{b} \, , 
\label{delta}
\end{eqnarray}
with identically null action on all other fields.  The label $N$ in $f^{abi(N)}$ means that the index $i$ belongs to the subgroup $N$. Also, distinct couplings $g_{N}$ are introduced for each subgroup and the coupling  $g^{\prime}$ is associated to the abelian subgroup along the symmetry breaking direction $z$. This concludes our summary of the BRST system of a SSB theory, more details can be obtained in \cite{ALVV22}.

Once the symmetry is broken, we can think of the gauge fixing as independent processes in each gauge direction. In the case of the unbroken directions, we may follow the procedure established in the last section, where we saw that the gauge  turns out to be fixed in the critical point of the functional ($\ref{Ol}$).

Our interest now rely on the broken directions. We can study the following generalized functional for them
\begin{eqnarray}
O_{BP}&=&  \int d^{4}x \left( \frac{1}{2} A^{a}_{\mu} A^{a}_{\mu}  + \frac{a_{1}}{2} \overline{c}^{a}c^{a}+ \frac{a_{2}}{2}\chi^{a}\chi^{a} \right) \, .
\label{OBP}
\end{eqnarray}
We calculate  the first variation of this functional as

\begin{eqnarray}
   \frac{\boldsymbol{\delta} (s_{q}+\delta)O_{BP}}{\boldsymbol{\delta} c^{a}}   = \left(\nabla_{\mu}A_{\mu}\right)^{a} + \frac{ia_{1}}{2} b^{a} - \frac{a_{1}}{2} \sum_{N} g_{N} f^{abi} \overline{c}^{b} c^{i} - \frac{a_{1}}{2} g^{\prime} f^{abz} \overline{c}^{b} c^{z} - a_{2} \mu f^{abz} \chi^{b} \, 
\label{1vobp}
\end{eqnarray}
where $(\nabla_{\mu}A_{\nu})^{a}$ is the covariant derivative defined in  \cite{ALVV22}

\begin{eqnarray}
(\nabla_{\mu}A_{\nu})^{a}&=&\partial_{\mu}A^{a}_{\nu}-\sum_{N} g_{N}f^{abi(N)}A^{i(N)}_{\mu}A^{b}_{\nu}-g^{\prime}f^{abz}A^{z}_{\mu}A^{b}_{\nu} \, .
\label{dercov1}
\end{eqnarray}

We can also obtain the gauge fixing action for the broken directions following the same reasoning from ($\ref{SgfOe}$), adapted to the broken case

\begin{eqnarray}
S_{gfBP}&=&  \int d^{4}x \left\{ i{b}^{a}\left(\frac{\boldsymbol{\delta} (s_{q}+\delta)O_{BP}}{\boldsymbol{\delta} c^{a}}\right) - \overline{c}^{a}(s_{q}+\delta)\left(\frac{\boldsymbol{\delta} \left(s_{q}+\delta\right)O_{BP} }{\boldsymbol{\delta} c^{a}}\right) \right\}  \, .
\label{SgfBP}
\end{eqnarray}

The sector of this $S_{gfBP}$ containing the Lagrange multiplier $b^{a}$ may be written in the form
\begin{eqnarray}
 \int d^{4}x \left(ib^{a}G^{a} -  \frac{a_{1}}{2}b^{a}b^{a}\right) \, .
\label{thgf}
\end{eqnarray}
with
\begin{eqnarray}
G^{a}= (\nabla_{\mu}A_{\mu})^{a}  -a_{1}\sum_{N} g_{N} f^{abi} \overline{c}^{b} c^{i} - a_{1}g^{\prime} f^{abz} \overline{c}^{b} c^{z}- a_{2}g\mu f^{abz}\chi^{b} \, .
\label{G}
\end{eqnarray}
In this way we can recognize this gauge fixing as one of the 't Hooft kind \cite{Hoo71, Fuj72} , designed in order to simultaneously show the unitarity and renormalizability of a theory with a SSB process. Also this gauge allows the elimination of non-physical couplings, and the usual choice demands $ a_{1} = -a_{2}  $ as being the 't Hooft gauge parameter, which we take as $\alpha$ from now on. This point deserves to be highlighted, the achievement of a 't Hooft gauge fixing action from a generating functional of the Morse kind. This is the development that we were looking for in order to deal with the Gribov problem in the broken phase of a SSB theory.

The equation of motion of the Lagrange multiplier then gives us the gauge condition for this gauge, and, as we learned from the example of the Feynman's gauge of the last section, we can fix the multiplier as

\begin{eqnarray}
b^{a}=   ig\mu f^{abz}\chi^{b} -i\sum_{N} g_{N} f^{abi} \overline{c}^{b} c^{i} - ig^{\prime} f^{abz} \overline{c}^{b} c^{z} \, ,
\label{bBP}
\end{eqnarray}
leading us to the gauge condition

\begin{eqnarray}
 (\nabla_{\mu}A_{\mu})^{a} =0 \, .
\label{daBP}
\end{eqnarray}
This is the equivalent to the Landau gauge condition here in this case of a SSB theory. This in fact is the result we obtain as the critical point of the Morse functional

\begin{eqnarray}
O_{M}&=&  \int d^{4}x \left( \frac{1}{2} A^{a}_{\mu} A^{a}_{\mu} \right)  \, ,
\label{OM}
\end{eqnarray}
under displacements generated by $s_{q}+\delta$. What we are saying is that the gauge fixing obtained in ($\ref{SgfBP}$), with the gauge conditions ($\ref{bBP}$) and ($\ref{daBP}$), actually fixes the gauge at the critical points of ($\ref{OM}$).

Now we can study the equation of motion of the anti-ghost $\overline{c}^{a}$ in this gauge fixing. In the Landau and Feynman cases we saw that this equation, when evaluated at the gauge configurations satisfying the gauge fixing condition (($\ref{ddO}$) and ($\ref{eagg}$) respectively), led us to the same null second order variation ($\ref{gribeq}$), which is the source of the Gribov problem.
Then, taking this equation of motion from $S_{gfBP}$ in (($\ref{SgfBP}$), and assuming the conditions ($\ref{bBP}$) and ($\ref{daBP}$) we arrive at

\begin{align}
 -\partial^{2}c^{a} - 2 \sum_{N} g_{N}f^{aib(N)}A^{i(N)}_{\mu} \partial_{\mu}c ^{b} - 2 g^{\prime}f^{azb}A^{z}_{\mu}\partial_{\mu}c ^{b} - g^{\prime2}f^{azb}f^{bzc(N)}A^{z}_{\mu}A^{z}_{\mu}c ^{c} -&\\
 - g^{\prime}\sum_{N} g_{N}\left(f^{aib(N)}f^{bzc}+f^{azb}f^{bjc(N)}\right)A^{i(N)}_{\mu}A^{z}_{\mu}c ^{c} 
  - \sum_{N} g_{N}^{2}f^{aib(N)}f^{bjc}A^{i(N)}_{\mu}A^{j(N)}_{\mu}c ^{c}  
  &= \alpha {\mu}^{2} f^{abz}f^{bcz} c^{c}\nonumber
\, .
\label{grBP}
\end{align}
This is the equivalent of ($\ref{gribeq}$) in this broken phase. More than this, one can show that the left hand side of this equation is just the second order variation of the Morse functional ($\ref{OM}$) at the critical point ($\ref{daBP}$),
\begin{eqnarray}
 (s_{q}+\delta)\left(\frac{\boldsymbol{\delta}  (s_{q}+\delta)O_{M} }{\boldsymbol{\delta} c^{a}}\right)  = \alpha {\mu}^{2} f^{abz}f^{bcz} c^{c}
\, .
\label{dercov2}
\end{eqnarray}
This becomes a surprise in relation to the usual Landau case. The vacuum expectation value of the scalar field induces  a contribution to the right hand side of the equivalent of the Gribov equation in a SSB theory. And in the case when we have a breaking with a definite sign for the algebraic combination $f^{abz}f^{bcz}$ in all possible broken directions, then the theory with the gauge fixing ($\ref{SgfBP}$), constructed from the generating functional ($\ref{OBP}$), will have its gauge fixed at the configuration corresponding to the minimum of the Morse functional ($\ref{OM}$), i.e., it will be localized in Gribov's first region. In this case, we say that the Gribov problem is solved along these broken directions.

\section{The Gribov Problem in $SU(2)\times U(1)$ with SSB}

In this Section we present how we can understand that there is not a Gribov problem in the $SU(2)\times U(1)$ gauge theory after a SSB nechanism. This is an instructive example, not only for the use of the concepts just developed, but mainly to answer an old doubt of how it is possible that the Gribov problem be solved in the eletroweak theory without demanding the confinement of its gauge bosons.
We begin by fixing the notation. Using the generators $(T^{A})_{ij}$ of the $SU(2)$, we define the gauge fields $A{\mu}^{ij}$ and ghosts $c_{ij}$
\begin{eqnarray}
A_{\mu}^{ij} &=& (T^{A})^{ij} A^{A}_{\mu} \, ,\nonumber \\
c^{ij}&=& (T^{A})^{ij} c^{A} \, .
\label{aij}
\end{eqnarray}
We also introduce the abelian field $ a_{\mu}$ and its ghost $q$, and the BRST transformations
\begin{eqnarray}
s A_{\mu}^{ij} &=& - \left(\partial_{\mu}c^{ij} -ig A_{\mu}^{il}c^{lj} + ig c^{il}A_{\mu}^{lj}\right)  \, , \nonumber \\
s c^{ij}&=& -ig c^{il}c^{lj}  \, , \nonumber \\
s a_{\mu} &=& -\partial_{\mu} q  \, ,  \nonumber \\
s c&=& 0 \, ,
\label{saij}
\end{eqnarray}
In order to simplify the presentation, we make the redefinitions  $B_{\mu}^{ij}= A_{\mu}^{ij} + \frac{e}{g} a_{\mu}\delta^{ij} $ and $Q_{ij} = c_{ij} + \frac{e}{g} q\delta_{ij}$, obtaining a condensed form for all the set of BRST transformations
\begin{eqnarray}
s B_{\mu}^{ij} &=& - \left(\partial_{\mu}Q^{ij} -ig B_{\mu}^{il}Q^{lj} + ig Q^{il}B_{\mu}^{lj}\right) \, ,  \nonumber \\
s Q_{ij} &=& -ig Q_{il}Q_{lj}\nonumber \, . \\
\end{eqnarray}

We also have scalar fields transforming as \footnote{The charges of the scalar and vector fields have been chosen equal for notation simplification.} 
\begin{eqnarray}
s \phi^{i} &=& -ig Q^{ij} \phi^{j}   \, ,\nonumber \\
s \phi^{\dagger i} &=& ig \phi^{\dagger j} Q^{ji}  \, ,
\label{sphii}
\end{eqnarray}
with the covariant derivatives
\begin{eqnarray}
(D_{\mu}\phi )^{i} &=& \partial_{\mu}\phi^{i} -ig B_{\mu}^{ij}  \phi^{j}  \, ,\nonumber \\
(D_{\mu}\phi )^{\dagger i} &=& \partial_{\mu}\phi^{\dagger i} +ig \phi^{\dagger j} B_{\mu}^{ji}  \, .
\label{dsu2}
\end{eqnarray}
Then we suppose that this theory undergoes a SSB with the scalar fields acquiring the vacuum expectation value $\mu$ 

\begin{eqnarray}
\phi_{i} \rightarrow \phi_{i} + \mu v_{i} \, , \nonumber \\
\phi^{\dagger}_{i} \rightarrow  \phi^{\dagger}_{i}+ \mu v_{i} \, ,
\label{vevsu2}
\end{eqnarray}
where  we opportunely define the following isovectors in internal space

\begin{equation}
\vec{v}\Rightarrow \left(
\begin{array}{c}
0\\
1
\end{array}
\right)\,\,\,\,
\vec{u}\Rightarrow \left(
\begin{array}{c}
1\\
0
\end{array}
\right) \, .
\label{vect}
\end{equation} 

At the broken phase, in the scalar sector of the theory the covariant derivatives (\ref{dsu2}) will generate masses for some of the gauge fields and the well known non-physical coupling. The quadratic terms dependent on $\mu$ are as follows

\begin{eqnarray}
\frac{1}{2}(D_{\mu}\phi)_{i}(D_{\mu}\phi)^{\dagger}_{i}& \rightarrow & \frac{g^{2}}{2}\mu^{2} v_{l}(B_{\mu})_{lj} (B_{\mu})_{ji}v_{i} - \frac{ig}{2}\mu \left(v_{l}\partial^{\mu}(B_{\mu})_{lj}\phi_{j} - \phi^{\dagger}_{l}\partial^{\mu}(B_{\mu})_{lj}v_{j}\right) \, .
\label{npc}
\end{eqnarray} 
The elimination of this last non-physical element in the eletroweak action is a major guide that we will use in the definition of the physical gauge of the 't Hooft kind.

At this moment, before we continue to the BRST analysis of the broken phase, we use the isovectors (\ref{vect}) to establish a projection of the gauge fields $B_{\mu}^{ij}$ onto the form  $W_{\mu}^{+}$, $W_{\mu}^{-}$, $Z_{\mu}$ e $\gamma_{\mu}$ as is usual in the literature

\begin{eqnarray}
u_{i}(B_{\mu})_{ij}u_{j} &=&  A_{\mu}^{3} + \frac{e}{g}a_{\mu} = \gamma_{\mu} \, , \nonumber \\
u_{i}(B_{\mu})_{ij}v_{j} &=& A_{\mu}^{1} -iA_{\mu}^{2} = W_{\mu}^{-} \, ,  \nonumber \\
v_{i}(B_{\mu})_{ij}u_{j} &=& A_{\mu}^{1} +iA_{\mu}^{2} = W_{\mu}^{+} \, ,  \nonumber \\
v_{i}(B_{\mu})_{ij}v_{j} &=& -A_{\mu}^{3} + \frac{e}{g}a_{\mu} = Z_{\mu} \, ,
\label{ZWG}
\end{eqnarray}
and the respective ghosts
\begin{eqnarray}
u_{i}(Q)_{ij}u_{j} &=&  c^{3} + \frac{e}{g}q_{\mu} =Q^{\gamma}   \, , \nonumber \\
 u_{i}(Q)_{ij}v_{j} &=& c^{1} -ic^{2} = Q^{-}  \,  , \nonumber \\
v_{i}(Q)_{ij}u_{j} &=& c^{1} +ic^{2} = Q^{+}  \, ,  \nonumber \\
   v_{i}(Q)_{ij}v_{j} &=& -c^{3} + \frac{e}{g}q_{\mu} = Q^{Z}  \,.
\end{eqnarray}
Then, it is useful to adopt the same redefinition for the scalars
\begin{eqnarray}
u_{i}\phi_{i}& =&\varphi_{1} \, ,  \hspace{1cm} v_{i}\phi_{i} =\varphi_{2} \, ,  \nonumber \\
\phi_{i}^{\dagger}u_{i}& =&\varphi_{1}^{\dagger} \, ,   \hspace{1cm} \phi_{i}^{\dagger}v_{i} =\varphi_{2}^{\dagger} \, .
\end{eqnarray}
If we notice that $\vec{v}$ e $\vec{u}$ satisfy $\delta_{ij} = v_{i}v_{j} + u_{i}u_{j}$ and $u_{i}v_{i}=0$, we realize from (\ref{npc}) that only the gauge field $\gamma_{\mu}$ associated to a residual  $U(1)$ symmetry will not develop a mass at the broken phase. Then we understand that this field can be associated to the direction $z$ and the others  $ Z_{\mu}$, $W_{\mu}^{-}$ and $ W_{\mu}^{+}$  to the directions $a$ in the notation introduced in Section 3. Then, the BRST operators at this phase will follow the general structure of ($\ref{sq}$)

\begin{align}
s_{q} \gamma_{\mu} =& -\partial_{\mu} Q^{\gamma} \, , &&\nonumber \\
s_{q} W_{\mu}^{+}=&  igW_{\mu}^{+}Q^{\gamma} \, ,& 
s_{q} W_{\mu}^{-}=& - igW_{\mu}^{-}Q^{\gamma} \, , \nonumber \\
s_{q} Q^{+} =& - ig Q^{+}Q^{\gamma} \, ,  &
s_{q} Q^{-} =& ig Q^{-}Q^{\gamma} \, , \nonumber \\
s_{q} \varphi_{1}=& -igQ^{\gamma}\varphi_{1} \, ,  &
s_{q} \varphi_{1}^{\dagger} = &igQ^{\gamma}\varphi_{1}^{\dagger} \, ,  \nonumber \\
s_{q} \overline{Q}^{z} =& i b^{z}\, ,&&  \nonumber \\
s_{q} \overline{Q}^{+} =& i b^{+} \, , &s_{q} \overline{Q}^{-} =& i b^{-} \, , \nonumber \\
\label{sqsu2}
\end{align}
and ($\ref{delta}$)

\begin{align}
\delta Z_{\mu}=& -\partial_{\mu}Q^{z} \, , \nonumber \\
\delta W_{\mu}^{+}=&- \left(\partial_\mu Q^{+} +ig \gamma_{\mu}Q^{+}  \right)= -(\nabla_{\mu}Q)^{+} \, , \nonumber \\
\delta W_{\mu}^{-}=& - \left(\partial_\mu Q^{-} -ig \gamma_{\mu}Q^{-} \right)=-(\nabla_{\mu}Q)^{-} \, , \nonumber \\
\delta \varphi_{1}=& -ig\mu Q^{-} \, , \hspace{1,5cm}  
\delta\varphi_{1}^{\dagger}= ig\mu Q^{+} \, , \nonumber \\
\delta \varphi_{2} =& -ig\mu Q^{z} \, ,
 \hspace{1,5cm}
\delta  \varphi_{2}^{\dagger}=ig\mu Q^{z} \, .
\label{deltasu2}
\end{align}
All other transformations not specified are identically null. We also introduced the anti-ghosts $\overline{Q}^{z}$, $\overline{Q}^{+}$, $\overline{Q}^{-}$, and respective Lagrange multipliers $ b^{z}$, $ b^{+}$ and $b^{-}$ in (\ref{sqsu2}). This structure assures that $s_q^2=\delta^2=\left\{s_q,\delta\right\}=0$.

Once we have defined the BRST structure of the broken phase, we can start the construction of the gauge fixing sector from the analogous of the generalized functional in (\ref{OBP}) translated to this $SU(2)XU(1)$ case

\begin{eqnarray}
O_{ew}&=& \int d^{4}x \left(  W_{\mu}^{+}W_{\mu}^{-} + \frac12 Z_{\mu} Z_{\mu} + \alpha (\varphi_{1}^{\dagger}\varphi_{1} + \varphi_{2}^{\dagger}\varphi_{2}) + \alpha^\prime (  \overline{Q}^{+} Q^{-}+ \overline{Q}^{-} Q^{+} + \overline{Q}^{z}Q^{z})\right) \, .
\label{Oew}
\end{eqnarray}
If we follow the same steps according to  (\ref{OBP}) we find

\begin{eqnarray}
\frac{\boldsymbol{\delta}( s_{q} + \delta) O_{ew}}{\boldsymbol{\delta}  Q^{+}} &=&  ( \nabla_{\mu}W_{\mu})^{-} -i\alpha^\prime g \overline{Q}^{-}Q^{\gamma} +i \alpha g\mu  \varphi_{1}  +i \alpha^\prime  b^{-} \equiv G^{-}+i\alpha^\prime b^{-} \, , \nonumber \\
\frac{\boldsymbol{\delta}( s_{q} + \delta) O_{ew}}{\boldsymbol{\delta}  Q^{-}} &=&  ( \nabla_{\mu}W_{\mu})^{+}   +i\alpha g \overline{Q}^{+}Q^{\gamma} -i \alpha g\mu  \varphi_{1}^\dagger +i \alpha^\prime  b^{+} \equiv G^{+}+i \alpha^\prime  b^{+} \, , \nonumber \\
\frac{\boldsymbol{\delta}( s_{q} + \delta) O_{ew}}{\boldsymbol{\delta}  Q^{z}} &=& \partial_{\mu}Z_{\mu}  + i \alpha  g\mu (\varphi_{2}- \varphi_{2}^{\dagger})  +i \alpha^\prime  b^{z} \equiv G^{z} +i \alpha^\prime  b^{z} \, .
\end{eqnarray}
We then use these results to obtain the gauge fixing action of the 't Hooft kind for this eletroweak theory

\begin{eqnarray}
S_{gfew}&=& ( s_{q} + \delta)  \int d^{4}x \left(\overline{Q}^{+} \left(G^{-}+i\alpha^\prime b^{-}\right) + \overline{Q}^{-} \left(G^{+}+i \alpha^\prime  b^{+}\right) + \overline{Q}^{z} \left(G^{z}+i \alpha^\prime  b^{z}\right)\right)
\label{SgfOew}
\end{eqnarray} 
As before, if we fix the multipliers as

\begin{eqnarray} 
2  b^{-} -  2g \overline{Q}^{-}Q_{\gamma} +  \frac{\alpha}{\alpha^\prime}g\mu  \varphi_{1} &=& 0\nonumber \\
2  b^{+}+  2g \overline{Q}^{+}Q_{\gamma} - \frac{\alpha}{\alpha^\prime}g\mu  \varphi_{1}^{\dagger} &=& 0 \nonumber \\
 2  b^{z} \frac{\alpha}{\alpha^\prime}g\mu \left(\varphi_{2}- \varphi_{2}^{\dagger}\right)&=&0.
 \label{bsu2}
\end{eqnarray}
The gauge conditions coming from the equations of motion of these multipliers lead us to
\begin{eqnarray}
  (\nabla_{\mu}W_{\mu})^{-}=0 \, , \hspace{1,5cm} (\nabla_{\mu}W_{\mu})^{+}=0 \, , \hspace{1,5cm} \partial_{\mu}Z_{\mu} = 0 \, .
\end{eqnarray}
These may be seen as the first order variation of the functional

\begin{eqnarray}
O_{Mew}&=& \int d^{4}x \left(  W_{\mu}^{+}W_{\mu}^{-} +\frac 12  Z_{\mu} Z_{\mu} \right)\, . 
\label{OMew}
\end{eqnarray}
This means that under the conditions (\ref{bsu2}), the gauge fixing (\ref{SgfOew}) obtained from (\ref{Oew}) is fixing the gauge on the critical point of $O_{Mew}$. Finally, we see that the Gribov conditions for this theory calculated as the equations of motion for the anti-ghosts $\overline{Q}^{+}$ , $ \overline{Q}^{-}$ and $\overline{Q}^{z}$  from (\ref{SgfOew}), after setting $\alpha=\alpha^\prime$, may be rewritten as

\begin{eqnarray}
-(s_{q}+\delta) \frac{\boldsymbol{\delta} (s_{q} + \delta)O_{M}}{\boldsymbol{\delta} Q^{+}}& =& \alpha  g^{2} \mu^{2} Q^{-} \, , \nonumber \\
-(s_{q}+\delta) \frac{\boldsymbol{\delta} (s_{q}+\delta)O_{M}}{\boldsymbol{\delta} Q^{-}}& =& \alpha  g^{2} \mu^{2} Q^{+} \, , \nonumber \\
-(s_{q}+\delta) \frac{\boldsymbol{\delta} (s_{q}+\delta)O_{M}}{\boldsymbol{\delta} Q^{z}}& =& 2\alpha  g^{2} \mu^{2} Q^{z} \, .
 \label{Gsu2}
\end{eqnarray}
The conclusion is that the gauge fixing action (\ref{SgfOew}) is actually fixing the gauge on the minimum of the Morse functional (\ref{OMew}). In other words, the gauge configurations allowed by this gauge fixing belong to Gribov's first region when we define the topography for the gauge configurations' manifold as the Morse functional (\ref{OMew}). This shows how the Gribov problem is solved in the broken phase of the eletroweak theory.

\section{The $SL(3,c)$ Complex Theory }

We now address this analysis to the $SL(3,c)$ complex gauge theory of $\cite{ALVV20}$, an environment where a SSB mechanism triggers the confinement of gauge bosons and fermions simultaneously. In its symmetric phase, the theory displays a $SL(3,c)$ gauge symmetry with a pair of a complex gauge field $\mathcal{A}_{\mu }^A$ and its conjugate  $\bar{\mathcal{A}}_{\mu }^A$. They couple to a pair of real scalar fields $\varphi^{A}$ and $\psi^{A}$ in an action given by

\begin{eqnarray}
 S= \int d^{4}x \left(\frac{i}{4}\mathcal{F}^{a}_{\mu\nu}\mathcal{F}^{a}_{\mu\nu}-\frac{i}{4}\bar{\mathcal{F}}^{a}_{\mu\nu}\bar{\mathcal{F}}^{a}_{\mu\nu} + Tr(\nabla_{\mu}\varphi)(\nabla_{\mu}\psi)  + V(\varphi,\psi )  \right)+ S_{GF} \, ,
 \label{action}
\end{eqnarray}
with $\mathcal{F}$ the curvature defined by

\begin{eqnarray}
\mathcal{F}_{\mu \nu}(\mathcal{A})=\partial_\mu \mathcal{A}_\nu - \partial_{\nu } \mathcal{A}_\mu -ig[\mathcal{A}_\mu, \mathcal{A}_\nu] \, ,
\label{F}
\end{eqnarray}
and ${\bar{\mathcal{F}}}$ its complex conjugate. 
The covariant derivatives when expressed in terms of commutators and anti-commutators are

\begin{eqnarray}
 \nabla_{\mu } \varphi &=& \partial_{\mu }\varphi +\frac{ig}{2} \left( -\{ \bar{\mathcal{A}}_{\mu }- \mathcal{A}_{\mu },\varphi \} - [\bar{\mathcal{A}}_{\mu }+ \mathcal{A}_{\mu },\varphi ] \right) \, , \nonumber \\
 \nabla_{\mu } \psi &=& \partial_{\mu }\psi +\frac{ig}{2} \left(\{ \bar{\mathcal{A}}_{\mu }- \mathcal{A}_{\mu },\psi \}-[\bar{\mathcal{A}}_{\mu }+ \mathcal{A}_{\mu },\psi ] \right) \, . \label{comnablapsi}
\end{eqnarray}
The symmetry breaking potential $ V(\varphi,\psi ) $

\begin{eqnarray}
 V(\varphi, \psi )= -\frac{m^{2}}{2} \varphi^{a} \psi^{a}+\frac{\lambda }{4}\left(\varphi^{a} \psi^{a}\right)^{2} \,  ,
 \label{potential}
\end{eqnarray}
has minima along the condition
\begin{equation}
  <\varphi^{a} \psi^{a}>= \frac{m^{2}}{\lambda } \, .
 \label{minima}
\end{equation}
In the broken phase, the scalar fields acquire vacuum expectation values 
\begin{eqnarray}
\varphi&\mapsto& \varphi + \mu \, , \nonumber \\
\psi&\mapsto& \psi + \mu  \, ,
\label{vevsl31}
\end{eqnarray}
and in the special case when 

\begin{eqnarray}
\mu &=& \frac{\wt\mu}{g}\left(\frac{\sqrt{2}}{\sqrt 3}T^{8} -\frac1{\sqrt3}T^{0}\right)\hspace{1cm} \, , \, \hspace{1cm} T^{0}=\frac{1}{\sqrt{6}}I \, ,
\label{vevsl32}
\end{eqnarray}
with
\begin{eqnarray}
\wt\mu &=&\frac{ mg}{\sqrt{\lambda}} \, ,
\label{a0}
\end{eqnarray}
this phase will develop gluon condensates \cite{ALVV20}.

After the SSB, as shown in \cite{ALVV23}, the independent cohomological classes are identified by the broken phase nilpotent BRST operator $s_q$

\begin{eqnarray}
s_q\mathcal{A}_{\mu }^i&=&-\left(\partial_{\mu }c^i+g_1f^{ijk}\mathcal{A}_{\mu }^jc^k\right) = -(D_{\mu}c)^{i}  \,  ,
\hspace{0.3cm} s_q\bar{\mathcal{A}}_{\mu }^i=-\left(\partial_{\mu }\bar{c}^i+g_1f^{ijk}\bar{\mathcal{A}}_{\mu }^j\bar{c}^k\right)= -(\overline{D_{\mu}c})^{i} \label{sqAi}  \,  , \nonumber \\
s_q\mathcal{A}_{\mu }^a&=&-\left(g_1f^{abi}\mathcal{A}_{\mu }^bc^i+g' f^{ab8}\mathcal{A}_{\mu }^bc_R^8  \right)  \,  ,
\hspace{1.1cm}  s_q\bar{\mathcal{A}}_{\mu }^a=-\left(g_1f^{abi}\bar{\mathcal{A}}_{\mu }^b\bar{c}^i+g'f^{ab8}\bar{\mathcal{A}}_{\mu }^bc_R^8  \right) \label{sqAa}  \,  , \nonumber \\
s_q \mathcal{A}^{8}_{R\mu} &=& -\partial_{\mu}c_R^8  \,  ,  \hspace{1,2cm}  s_q \mathcal{A}^{8}_{I\mu} = 0 \label{sqA8}  \,  , \nonumber \\
s_q{ {c}}^i &=& \frac{g_1}{2}f^{ijk}{c}^j{c}^k  \,  ,  \hspace{1cm} s_q{ \bar{c}}^i = \frac{g_1}{2}f^{ijk}\bar{c}^j\bar{c}^k  \,  , \hspace{1cm} s_q c_R^8=0  \,  , \hspace{1cm} \nonumber \\
s_{q}\varphi^{a} &=& \frac{ig_{1}}{2}\left(d^{abj}\varphi^{b}(c-\bar{c})^{j}+ i f^{abi}\varphi^{b}(c+\bar{c})^{i}\right) -g'f^{ab8}\varphi^{b}c_{R}^{8}  \,  , \nonumber \\
s_{q}\psi^{a} &=& \frac{ig_{1}}{2}\left(-d^{abj}\psi^{b}(c-\bar{c})^{j}+ i f^{abi}\psi^{b}(c+\bar{c})^{i}\right) -g'f^{ab8}\psi^{b}c_{R}^{8}  \,  ,  \nonumber \\
s_{q} c^{a}&=& g_{1} f^{abi}c^{b}c^{i} + g'f^{ab8}c^{b}c_{R}^{8}  \,  , \hspace{1,5cm}
s_{q}\bar{c}^{a}= g_{1} f^{abi}\bar{c}^{b}\bar{c}^{i} + g'f^{ab8}\bar{c}^{b}c_{R}^{8}  \,  , \nonumber \\
s_{q} q^{A} &=& - i b^{A}  \,  , \hspace{1cm} s_{q}b^{A}=0  \,  , \nonumber \\
s_{q} \bar{q}^{A} &=& i \bar{b}^{A}  \,  ,  \hspace{1.3cm} s_{q}\bar{b}^{A}=0  \,  , \hspace{1cm} b^{A} \rightarrow (b^{i}, b^{a},b^{8})  \,  .
\label{sqc}
\end{eqnarray}

The new coupling constant $g_1$ is associated to the the residual  $SL(2,C)$ symmetry of the new vacuum. Also, from (\ref{sqA8}), we recognize an abelian symmetry with coupling $g'$ associated to the real part $\mathcal{A}^{8}_{R\mu}$ of the complex gauge component $\mathcal{A}^{8}_{\mu}$. This expression also shows that the imaginary part $\mathcal{A}^{8}_{I\mu}$ now becomes a vectorial matter field. This happens because the imaginary component $c_I^8$ ceases to be a ghost of the BRST operator  $s_q$, and only the real component $c_R^8$ appears as an abelian ghost after the phase transition. This frame is actually responsible for the development of a confining fermionic potential in the asymmetric phase of the complex theory \cite{ALVV20}. Equations (\ref{sqAa}) feature another fundamental outcome of the phase transition: they sign the braking of the holomorphicity of the complex theory. This has an amazing impact on the physics of this theory, and is ultimately responsible for the emergence of gluon condensates as composite particles \cite{ALVV23} . 

After this brief resume of the previous results on the $SL(3,c)$ complex gauge theory, we may now proceed to the application of the general analysis established in Section 3 to this specific case. For this, we start by writing the analogous of the delta operator in (\ref{delta})

\begin{eqnarray}
\delta \mathcal{A}_{\mu }^{a}&=& - \left(\partial_{\mu}c^{a}+g_{1} f^{aib} \mathcal{A}_{\mu }^{i}c^{b} + g'f^{a8b}\mathcal{A}_{R \mu}^{8}c^{b}\right) \equiv -(\nabla_{\mu}c)^{a}  \,  , \nonumber \\
\delta \bar{\mathcal{A}_{\mu }}^{a}&=& - \left(\partial_{\mu}\bar{c}^{a}+g_{1} f^{aib} \bar{\mathcal{A}_{\mu }}^{i}\bar{c}^{b} + g'f^{a8b}\bar{\mathcal{A}}_{R \mu}^{8}\bar{c}^{b}\right) \equiv -(\bar{\nabla}_{\mu}\overline{c})^{a}  \,  , \nonumber \\
\delta \varphi^{a} &=&-i\frac{\wt\mu }{3\sqrt{2}}(c^{a}-\bar{c}^{a}) + i\frac{\wt\mu }{\sqrt{6}}d^{a8c}(c^{c}-\bar{c}^{c})-  \frac{\wt\mu }{\sqrt{6}}f^{a8c}(c^{c}+\bar{c}^{c})  \,  , \nonumber \\
\delta \psi^{a} &=& i\frac{\wt\mu }{3\sqrt{2}}(c^{a}-\bar{c}^{a}) -i \frac{\wt\mu }{\sqrt{6}}d^{a8c}(c^{c}-\bar{c}^{c}) -  \frac{\wt\mu }{\sqrt{6}}f^{a8c}(c^{c}+\bar{c}^{c})  \,  .
\label{deltac}
\end{eqnarray}

This choice assures  that $s_{q}+\delta$ becomes nilpotent. This is a  requirement in order that this combined operator be used in the construction of the gauge fixing sector $S_{GF}$ of the action (\ref{action}) in the broken phase, i.e., a gauge fixing of the 't Hooft kind. We must remember that, as before, we are implicit assuming that there is a conventional gauge fixing along the directions that remain symmetric, and our attention is now exclusively on the broken directions. It is also convenient to show that in the special case of the breaking described in (\ref{vevsl32}) the following expression applies

\begin{equation}
d^{ab8}= -\frac{1}{2\sqrt{3}}\delta^{ab} \,  .
\label{dab8}
\end{equation} 

Following the procedure of Section 3, we define the functional analogous of (\ref{OBP}) adapted to the present case

\begin{eqnarray}
O&=& \int d^{4}x \left(\frac{i}{2}\mathcal{A}_{\mu}^{a}\mathcal{A}_{\mu}^{a}-\frac{i}{2}\bar{\mathcal{A}}_{\mu}^{a}\bar{\mathcal{A}}_{\mu}^{a} +  a_{2} \varphi^{a}\psi^{a} + a_{1}(q^{a}c^{a} - \bar{q}^{a}\bar{c}^{a} ) \right)  \,  .
\label{O}
\end{eqnarray}
Using (\ref{sqc}), ( \ref{deltac}) and (\ref{dab8}), the first variation of this functional is

\begin{eqnarray}
(s_{q}+\delta) O&=&  \int d^{4}x \left\{\left( i c^{a}(\nabla_{\mu}\mathcal{A}^{\mu})^{a} - i\bar{c}^{a}(\bar{\nabla}_{\mu}\bar{\mathcal{A}}^{\mu})^{a}) -i a_{1}(\bar{b}^{a}\bar{c}^{a} +b^{a}c^{a}\right)\right.\nonumber \\
&&\left.- a_{1}\left(g_{1}f^{abi} q^{a}c^{b}c^{i} + g'f^{ab8}q^{a}c^{b}c_{R}^{8}-g_{1}f^{abi}\bar{q}^{a}\bar{c}^{b}\bar{c}^{i}-g'f^{ab8}\bar{q}^{a}\bar{c}^{b}c_{R}^{8}\right)\right.\nonumber \\
&&\left.+a_{2}\left(i \frac{\sqrt 2}{4}\wt\mu (c^{a}-\bar{c}^{a})(\varphi^{a}-\psi^{a}) - \frac{1}{\sqrt{6}}\wt\mu f^{ab8}(c^{a}+\bar{c}^{a})(\varphi^{b}+\psi^{b})\right)\right\} \,  ,
\end{eqnarray}
and then

\begin{align}
\frac{\boldsymbol{\delta} (s_{q}+\delta )O}{\boldsymbol{\delta} c^{a}} =&\,i (\nabla_{\mu}\mathcal{A}^{\mu})^{a} - i  a_{1}b^{a} - a_{1}g_{1}f^{abi}q^{b}c^{i} - a_{1}g'f^{ab8}q^{b}c_{R}^{8}\nonumber \\
&+ i\frac{\wt\mu }{2\sqrt 2}a_{2}(\varphi^{a}-\psi^{a}) -\frac {\wt\mu}{\sqrt 6} a_{2}f^{ab8}(\varphi^{b}+\psi^{b}) \equiv G^{a}- i  a_{1}b^{a}  \,  ,
\end{align}

\begin{align}
\frac{\boldsymbol{\delta} (s_{q}+\delta )O}{\boldsymbol{\delta} \bar{c}^{a}} =&i(\bar{\nabla}_{\mu}\bar{\mathcal{A}}^{\mu})^{a} -i a_{1} \bar{b}^{a}
+ a_{1}g_{1}f^{abi}\bar{q}^{b}\bar{c}^{i} + a_{1}g'f^{ab8}\bar{q}^{b}c_{R}^{8}\nonumber \\
&- i\frac{\wt\mu }{2\sqrt 2}a_{2}(\varphi^{a}-\psi^{a}) -\frac {\wt\mu}{\sqrt 6} a_{2}f^{ab8}(\varphi^{b}+\psi^{b}) \equiv \bar G^{a}- i  a_{1}b^{a}  \,  ,  .
\end{align}

With these results we can define the adapted version of the gauge fixing of a broken phase given in (\ref{SgfBP}) to this complex gauge theory yielding
\begin{eqnarray}
S_{GF}&=& (s_{q}+\delta) \int d^{4}x \left(q^{a}\frac{\boldsymbol{\delta} (s_{q}+\delta )O}{\boldsymbol{\delta} c^{a}} + \bar{q}^{a}\frac{\boldsymbol{\delta} (s_{q}+\delta )O}{\boldsymbol{\delta} \bar{c}^{a}} \right)\nonumber \\
&=& \int d^{4}x \left\{- ib^{a}( G^{a} -i a_{1}b^{a})+i \bar{b}^{a}(\bar{G}^{a} +i a_{1} \bar{b}^{a})-q^{a}(s_{q}+\delta) \frac{\boldsymbol{\delta} (s_{q}+\delta )O}{\boldsymbol{\delta} c^{a}}\right.\nonumber \\
&&\left.\hskip1.5cm - \bar{q}^{a}(s_{q}+\delta) \frac{\boldsymbol{\delta} (s_{q}+\delta )O}{\boldsymbol{\delta} \bar{c}^{a}}\right\} \,  .
\label{SGFSL3C}
\end{eqnarray}
This gauge fixing restores the unitarity of the action (\ref{action}) in the broken phase when the vacuum of (\ref{vevsl31}), (\ref{vevsl32}) and (\ref{a0}) is chosen. For this, the relation $a_{1}=-ia_{2}$ is demanded and we will use it in the next expressions.

The gauge conditions are then obtained in an analogous way as done in (\ref{bBP}) and (\ref{daBP}), requiring

\begin{align}
-&2b^{a} + 2ig_{1}f^{abi}q^{b}c^{i} + 2ig'f^{ab8}q^{b}c^{8}_{R} +i\frac{\sqrt2}{4}\wt\mu (\psi^{a}-\varphi^{a})+\frac{1}{\sqrt6} \wt\mu f^{ac8}(\psi^{c}+\varphi^{c})=0  \,  ,\\
-&2\bar{b}^{a} -2ig_{1}f^{abi}\bar{q}^{b}\bar{c}^{i} -2ig'f^{ab8}\bar{q}^{b}c^{8}_{R} +i\frac{\sqrt2}{4}\wt\mu (\psi^{a}-\varphi^{a})+\frac{1}{\sqrt6} \wt\mu f^{ac8}(\psi^{c}+\varphi^{c})=0  \,  ,
\end{align}  
so that

\begin{eqnarray}
(\nabla_{\mu}A^{\mu})^{a}&=&0. \nonumber \\
(\bar{\nabla}_{\mu}\bar{A}^{\mu})^{a}&=&0  \,  . 
\label{LGSL3C}
\end{eqnarray}

These last expressions are the equivalent of the Landau gauge condition for the broken directions in the complex theory. They are obviously associated to a critical point condition for the complex Morse functional

\begin{eqnarray}
O_1&=& \int d^{4}x \left(\frac{i}{2}\mathcal{A}_{\mu}^{a}\mathcal{A}_{\mu}^{a}-\frac{i}{2}\bar{\mathcal{A}}_{\mu}^{a}\bar{\mathcal{A}}_{\mu}^{a}  \right)  \,  .
\label{O1}
\end{eqnarray}

Now, assuming a Landau gauge fixing along the unbroken directions $i$ and $8$, we obtain equations of the Gribov type for the ghosts $c^{a}$ and $\bar{c}^{a}$ from the equations of motion of the anti-ghosts $q^{a}$ and $\bar{q}^{a}$ respectively

\begin{eqnarray}
-\frac{\boldsymbol{\delta} S_{GF}}{\boldsymbol{\delta} {q}^{a}}|_{(\nabla_{\mu}\mathcal{A}^{\mu})^{a}=0}&=& - (\nabla_{\mu}\nabla^{\mu}c)^{a} +\frac{\alpha}{4}\wt\mu^2 c^{a}=0 \,  , \nonumber \\
-\frac{\boldsymbol{\delta} S_{GF}}{\boldsymbol{\delta} \bar q^{a}}|_{(\bar{\nabla}_{\mu}\bar{\mathcal{A}}^{\mu})^{a}=0}&=&  (\bar{\nabla}_{\mu}\bar{\nabla}^{\mu}\bar{c})^{a} +\frac{\alpha}{4}\wt\mu^2 \bar{c}^{a}=0 \,  .
\label{GSL3C}
\end{eqnarray}
Here we have set $2a_1=i\alpha$ according to the language in \cite{ALVV23}.

Finally, these equations (\ref{GSL3C}) may be interpreted as local minimum conditions for the Morse functional (\ref{O1}) at the critical points determined by (\ref{LGSL3C}).

\section{Conclusion}

In this work we searched for an answer of why Gribov's problem for non-abelian gauge theories may be solved in practice in situations as different as the eletroweak theory, with its non-abelian $SU(2)$ symmetry, or at the same time in the case of the confining gluon theory. Our understanding is that the possible explanation would be associated to the SSB mechanism. In the eletroweak example, this mechanism is already well known, and in Section 4 we succeeded in applying the general structure proposed in Section 3, where we approached this problem by the point of view of Morse theory. There we showed how the construction of a gauge fixing action of the 't Hooft kind may be built from a Morse functional principle, and afterwards how a Gribov condition can be derived from such development. In the case of the eletroweak theory, this reasoning led to the expressions (\ref{Gsu2}), allowing an interpretation that the gauge configurations are limited to Gribov's first region once they are restricted  by the gauge fixing constructed from the Morse functional (\ref{Oew}). 

Then we applied the same treatment for a prototype theory with gluon confinement. The $SL(3,c)$ theory that we proposed in \cite{ALVV20} becomes a natural choice for this research, as it has a confined phase after a SSB process. Again starting from the functional (\ref{O}) we derived the gauge fixing action (\ref{SGFSL3C}) of the 't Hooft type for this phase of the complex theory, and this enabled us to show how this gauge fixes the gauge configurations as minima of the Morse functional (\ref{O1}) as expressed by the Gribov conditions (\ref{GSL3C}).

This point deserves to be reinforced. The broken phase of the $SL(3,c)$ theory confining fermions and gluons is achieved at the vacuum defined by (\ref{vevsl32}). As explained in \cite{ALVV20}, this option is in fact mandatory, as no other vacuum choice gives us the necessary characteristics to provide such confinement. Now, in the present work, we see that among all possible vacua for the broken phase, the same choice (\ref{vevsl32}) is just the one appropriate to lead to the Gribov conditions (\ref{GSL3C}), as it is the one which provides, in the general expression (\ref{dercov2}), a definite sign for the algebraic combination $f^{abz}f^{bcz}$ in all possible broken directions. Other possible breaking directions would not allow the gauge fixing described in Section 5, and then would not fix the gauge configurations in Gribov's first region. The fact that it is the same and unique vacuum that accomplishes both confinement and solves Gribov's problem simultaneously at this theory is remarkable. Some cautionary words should, however, be stressed as regards to the implication of these findings to the QCD model. There is still a theoretical gap in transposing the sl(3,C) framework here addressed to the su(3) symmetric QCD  case. 

As a final statement, we would like to point out that such conclusions may be interpreted as  arguments in favour of the comprehension of confinement as a phase of a theory after a SSB mechanism yet to be fully understood. However a complete study of the remaining gauge copies within the Gribov horizon, the fundamental modular region issue, is still lacking. Should it be possible to recast this problem in terms of Morse theory?


\begin{thebibliography}{10}
  
\bibitem{ALVV20}
R.~L.~P.~G.~Amaral, V.~E.~R.~Lemes, O.~S.~Ventura and L.~C.~Q.~Vilar,
``A Path to confine gluons and fermions through complex gauge theory,''
Phys. Rev. D \textbf{101} (2020) no.9, 094002
doi:10.1103/PhysRevD.101.094002
[arXiv:2002.07222 [hep-th]].


\bibitem{Gribov}
V.~N.~Gribov,
``Quantization of Nonabelian Gauge Theories,''
Nucl. Phys. B \textbf{139} (1978), 1
doi:10.1016/0550-3213(78)90175-X.


\bibitem{Milnor}
J. Milnor, Ann. Math. Stud. 51, Princeton Univ. Press, Princeton NJ
(1973).

\bibitem{Matsumoto}
Matsumoto, Yukio, “An Introduction to Morse Theory”, Iwanami series
in modern mathematics 208, AMS Bookstore (2002).

\bibitem{Baal}
P. van Baal, 
``More (thoughts on) Gribov copies'',
Nucl. Phys. B \textbf{369} (1992), 259.

\bibitem{LLVV14}
R.~R.~Landim, V.~E.~R.~Lemes, O.~S.~Ventura and L.~C.~Q.~Vilar,
``Revisiting Gribov’s Copies Inside The Horizon'' Eur. Phys. J. C \textbf{74} (2014) 11, 3069
[arXiv:1402.6235 [hep-th]].

\bibitem{Sor95}
O.~Piguet and S.~P.~Sorella,
``Algebraic renormalization: Perturbative renormalization, symmetries and anomalies,''
Lect. Notes Phys. Monogr. \textbf{28} (1995), 1-134
doi:10.1007/978-3-540-49192-7.



\bibitem{Zw12}
N.~Vandersickel and D.~Zwanziger,
``The Gribov problem and QCD dynamics,''
Phys. Rept. \textbf{520} (2012), 175-251
doi:10.1016/j.physrep.2012.07.003
[arXiv:1202.1491 [hep-th]].

\bibitem{CF}
G. Curci and R. Ferrari, 
``On a class of Lagrangian models for massive and massless Yang-Mills fields,''
 Nuovo Cim. A\textbf{32} (1976) 151.


\bibitem{ALVV22}
R.~L.~P.~G.~Amaral, V.~E.~R.~Lemes, O.~S.~Ventura and L.~C.~Q.~Vilar,
``BRST view of spontaneous symmetry breaking,''
Phys. Rev. D \textbf{105} (2022) no.12, 125007
doi:10.1103/PhysRevD.105.125007
[arXiv:2205.02903 [hep-th]].


\bibitem{Wei96}  S. Weinberg, “The Quantum Theory of Fields Vol. 2: Modern applications,” (Cambridge: Cambridge University Press (1996)).



\bibitem{Hoo71}G. 't Hooft, "Renormalizable Lagrangians for Massive Yang-Mills Fields,"
Nucl.Phys.B \textbf{35} (1971) 167.

\bibitem{Fuj72} K. Fujikawa, B. Lee and A. Sanda,
"Generalized Renormalizable Gauge Formulation of Spontaneously Broken Gauge Theories,"
Phys. Rev. D \textbf{6}, 2923 (1972).



\bibitem{ALVV23}
R.~L.~P.~G.~Amaral, V.~E.~R.~Lemes, O.~S.~Ventura and L.~C.~Q.~Vilar,
"BRST characterization of the broken phase observables of a confining complex theory,"
Phys. Rev. D \textbf{107} (2023) no.11, 115015
doi:10.1103/PhysRevD.107.115015
[arXiv:2304.04560 [hep-th]].










\end{thebibliography}
\end{document}